\documentstyle[aps,epsfig,float,manuscript]{revtex}
\draft

\begin{document}
\title{Site-diluted three dimensional Ising Model\\
        with long-range correlated disorder}
\author{
H.~G.~Ballesteros}
\address{
{\it Departamento de F\'{\i}sica Te\'orica I },\\ 
{\it Universidad Complutense de Madrid, 28040 Madrid, Spain,} \\
{\tt \small hector@lattice.fis.ucm.es.}} 
\author{and \\
 G.~Parisi.}
\address{
{\it Dipartimento di Fisica, 
Universit\`a di Roma I and INFN},\\ 
{\it P. A. Moro 2, 00185 Roma, Italy,}\\
{\tt \small giorgio.parisi@roma1.infn.it.}}

\date{March 15, 1999}
\maketitle

\thispagestyle{empty}

\begin{abstract}
We study two different versions of the site-diluted Ising model in three
dimensions with long-range spatially correlated disorder by Monte Carlo
means. We use finite-size scaling techniques to compute the
critical exponents of these systems, taking into account the strong 
scaling-corrections. We find a $\nu$ value that is compatible with 
the analytical predictions.
\end{abstract}


\section{Introduction}

Neutron and X-Ray critical scattering experiments in different systems
~\cite{EXPERIMENTS} revealed an unexpected feature: the two length
scales coexistence.  In the theory of  critical phenomena it is expected
that
the vector dependence of the scattering intensity corresponds to a
Lorentzian function with a width proportional to the inverse of the
correlation length. Nevertheless it was found~\cite{EXPERIMENTS} more
reasonable to understand the experimental measures supposing the
superposition of a broad Lorentzian (which width closely behaves as it
is theorically expected for these materials) and a sharper function, 
which behaviour is similar to a simple or a squared Lorentzian function. 
In Ref.~\cite{ALTARELLI} it is proposed that this new component of the
scattering intensity has an defect related origin, in particular due
to the presence of dislocations near the sample surface.  So there is
a crossover between the bulk critical behaviour (broad component) and
the {\it disorder} critical behaviour. As the defects are not randomly
placed points but randomly oriented lines, the quenched
disorder is long-range correlated.

A Gaussian disorder with correlations decaying like a power law was
studied in Ref.~\cite{WEINRIB} for the vector spin models by
analytical means, in particular using renormalization group expansion
in $\epsilon=4-d$ and $\delta=4-a$, up to first order, where $a$ is
the power of the potential decay of the spatial correlation function
and $d$ is the ordinary spatial dimension.  For example, straight
dislocation lines with random orientation can be represented with
$a=d-1$.  In addition to the Gaussian fixed point, the pure Ising point 
and the short-ranged disordered one it appears
another fixed point. This new point has a pair of complex eigenvalues
that lead to oscillating scaling corrections.  The critical exponents
of the long-range correlated disorder fixed point are
\begin{eqnarray}
\nu&=&\frac{2}{a}, \nonumber \\ \eta&=&{\cal O}(\epsilon^2),
\label{EXPOTHEORY} 
\end{eqnarray} 
being, as usual, $\nu$ the thermal critical exponent, associated to
the correlation length and $\eta$, the anomalous dimension of the
field.  Through the scaling relations it is possible to obtain the
other critical exponents of the system.  The first of the relations in 
Eq.~(\ref{EXPOTHEORY}) should be valid at all order in perturbation
theory and it should be an exact relation. The more interesting case
of non-Gaussian disorder has never been studied in detail, although it
is possible that the results for non-Gaussian disorder are the same of
the Gaussian case. It should be noted that the previous relation is valid 
only if the disorder decays in a sufficiently slow manner.  

A useful criterion due to Harris~\cite{HARRIS} tells us whether the
disorder is an irrelevant perturbation in the pure system, in terms of
the pure critical behaviour.  The criterion states that the
short-ranged disorder is not relevant when $d\nu_{\rm
pure}-2=-\alpha_{\rm pure}>0$.  In Ref.~\cite{WEINRIB} the Harris
criterion is extended to the long-range correlated case.  This kind of
disorder is an irrelevant perturbation if the condition $2/a >\nu_{\rm
pure}$ holds, i.e.  when the exponent $\nu$ given by the relation
(\ref{EXPOTHEORY}) is larger than the exponent $\nu_{\rm pure}$ of the
model without disorder.

A line of defects corresponds to a correlation decay with a $a=2$
power, but with a non-Gaussian distribution of the disorder.  In this
case, using the data from the pure Ising case,
$1>\nu=0.6294(5)(5)$~\cite{ISINGPERC}, the extended criterion tell us
that the disorder is a relevant perturbation.  Neglecting the
non-Gaussian effects and applying the results of the
Eq.~(\ref{EXPOTHEORY}) to this case, in Ref.~\cite{ALTARELLI}, is
found a $\nu$ exponent close with the experimental data for different
materials~\cite{EXPERIMENTS}.  Furthermore, it was computed the local
fluctuation on the critical temperature due to the line defects
concentration finding also an acceptable agreement.

In Ref.~\cite{PAPOULAR} the same authors have studied the influence of
the long-range correlated disorder in the line shape of the narrow
component of the scattering intensity function, finding that it can be
steeper than a Lorentzian one, also in agreement with the experimental
data. They were assuming the Gaussian disordered model and the line
defects one, where the disorder is non-Gaussian at large scale, do
belong to the same Universality Class.

In this study we will compute by Monte Carlo means the critical
exponents of the site-diluted Ising model with long-range correlated
disorder, in the $a=2$ case, in order to numerically check the
analytical predictions that seem compatible with some experimental
results. In order to do this we will use the finite-size scaling
techniques that are recently applied to the study of random
site-diluted Ising systems~\cite{ISDIL3D,ISDIL2D4D}.  We will study
both Gaussian and non-Gaussian disorder and we will find similar
results for the critical exponents supporting the correctness of the
analysis of Ref.~\cite{ALTARELLI,PAPOULAR}.

The layout of the paper is the following. In section 2 we will define
the model we have simulated in the lattice and the two different ways
used to introduce the long-range correlated disorder in the system.
In section 3 we will show the finite-size techniques we used.  The
technical details of the simulation will be reported in section 4. The
numerical results will be shown in section 5. Finally, in section 6,
we will report the conclusions of this study.

\section{The models and the observables}

We have considered the following Hamiltonian defined in a 
cubic lattice of linear size $L$ with periodic boundary conditions:
\begin{equation}
H=-\beta\sum_{<i,j>} \epsilon_i \epsilon_j \sigma_i \sigma_j\ ,
\label{HAMILTONIAN}
\end{equation}
where the sum is extended to the nearest neighbours , $\sigma$ are the
usual Z$_2$ spin variables and the $\epsilon$'s are quenched random
variables, with long-range spatial correlation.  An actual
${\epsilon_i}$ set will be called a {\it sample} from now on.  We have
studied two different ways to introduce the correlation between the
${\epsilon_i}$ variables.

The first one is to obtain a set of $V=L^3$ correlated Gaussian random 
variables, $\eta(\bbox x)$, where
$\bbox x$ is the position vector in the lattice, 
with these properties:
\begin{eqnarray}
\langle\eta(\bbox x)\rangle &=&0, \nonumber \\
\langle\eta(\bbox x) \, \eta(\bbox y) \rangle &\propto& 
\frac{1}{{|{\bbox {x-y}}|}^a} \ \ (\equiv C(|{\bbox {x- y}}|) ), 
\label{RANDOMCORREL}
\end{eqnarray}
where $d>a>0$. In order to do this we have used the Fourier Filtering (FF)
method~\cite{FFMETHOD}.

Let $\tilde C(\bbox p)$ be the Fourier transform of the function
$C(|{\bbox {x-y}}|)$ in momentum space. Let us define the set 
$\tilde \eta(\bbox p)$ as
\begin{equation}
\tilde \eta(\bbox p)= \sqrt{\tilde C(\bbox p)} \, u(\bbox p),
\end{equation}
where $u(\bbox p)$ is a Gaussian set of random numbers in the complex plane
with the following properties:
\begin{eqnarray}
\langle u (\bbox p)\rangle &=&0, \nonumber \\
\langle u(\bbox p) u(\bbox p) \rangle &=&1, \\
\langle u(\bbox p) u(\bbox p') \rangle&=& 0, \ 
\mbox{if} \ p\neq p'. \nonumber
\end{eqnarray}

At this point we construct $\eta(\bbox x)$ as the inverse Fourier
transform of the $\tilde \eta(\bbox p)$ set. In order to assure that
the $\eta(\bbox x)$ set is real we have to introduce the condition:
\begin{equation} 
\tilde \eta({\bbox {-p}})= \tilde \eta^*(\bbox p).
\label{REAL} 
\end{equation} 
The zero mode divergence in a lattice
treatment is eliminated by using the condition $u({\bbox
p=0})=0$. This choice agrees with the property $\langle u (\bbox
p)\rangle =0$. 

With these definitions $\eta(\bbox x)$ becomes a Gaussian random
variable, as it is a sum of a large number of random variables, and it
is easy to prove that the relations given by Eq.~(\ref{RANDOMCORREL})
are satisfied. Furthermore, it is also possible to calculate the variance 
of this Gaussian distribution, as the zero momentum inverse Fourier
transform of $\tilde C(\bbox p)$.

With the $\{\eta(\bbox x)\}$ set we proceed to choose, with a given
probability $p$, that we will call the mean concentration of the
system, when each site is occupied ($\epsilon_i=1$) or not
($\epsilon_i=0$).  We compute the area below the corresponding
Gaussian distribution for the $\eta$ random variable up to its actual
value $\eta(\bbox x)$. We compare this quantity with $p$ and we
consider that the site is occupied if the area is smaller than
$p$.

In this work we will study the case $a=2$ corresponding to linear
defects.  With this value we have checked that the correlation
obtained for the $\{\eta(\bbox x)\}$ set with the FF method, performed
in double precision, is in good agreement with the expected
correlation function. Although it is evident that the $\epsilon_i$ are
not Gaussian variables, their connected correlation functions at all
different points are equal to zero and therefore this model corresponds
to a Gaussian model at large scale (non-Gaussian effects are
restricted at short scale and are likely irrelevant). This model
is what we referred to as Gaussian distributed noise in the
introduction.

A second way to obtain samples with long-range correlated disorder
with a decay with the inverse of the square of the distance is to
remove lines of a given configuration.  We start with a filled cubic
lattice and remove lines until we get the fixed concentration $p$.
The last line considered in this procedure is removed or not with a
given probability in order to get as the mean concentration the $p$
value.  We also want that the probability of removal for all the
lattice points to be the same, and the lattice symmetries to be
preserved. We can do that by only removing lines along the axes.  It
is clear that the connected correlation functions with this method are
definitely different from zero also at long distances.  The noise is
very far from being Gaussian and this model is referred to as the
non-Gaussian distributed noise in the introduction.

In both cases, we will consider the quenched disorder, that is, we
first calculate the average of a given observable on the
$\{\sigma_i\}$ variables with the Boltzmann weight given by the
Hamiltonian of the Eq.~(\ref{HAMILTONIAN}), the results on the
different samples being {\it later} averaged. The quenched
approximation is due to the fact that the defect dynamics is slower
that the associated to the magnetic interaction.  We will denote by
brackets the thermal average and by overlines the sample average.  The
observables will be denoted with calligraphic letters, i.e.  $\cal O$,
and we will use the italics for the double average
$O=\overline{\langle{\cal O}\rangle}$.

Thus, we can define the nearest-neighbours energy as
\begin{equation}
{\cal E} =\sum_{\langle
i,j\rangle}\epsilon_i\sigma_i\epsilon_j\sigma_j\ .
\end{equation}
This quantity is extensively used for extrapolating the results 
for a given observable, $O$, obtained at coupling $\beta$ to a 
nearby coupling $\beta'$~\cite{FALCIONI} as well as for calculating 
$\beta$-derivatives through its connected correlation with the 
observable. For instance, one can define the specific-heat as
\begin{equation}
C \, =  \partial_\beta \overline{<{\cal E}>} \,=
\frac{1}{V}\left( \, \overline{\langle{\cal E}^2\rangle-{\langle{\cal
E}\rangle}^2} \, \right).
\end{equation}

The order parameter of the phase transition is the usual 
normalized magnetization
\begin{equation}
{\cal M}=\frac{1}{V}\sum_i \epsilon_i\sigma_i\ .
\end{equation}

As in a finite lattice, its mean value, $M$, is zero, we are restricted 
to work with even powers of the magnetization. The second power is 
related to the susceptibility of the system:
\begin{equation}
\chi=V\overline{\left\langle {\cal M}^2 \right\rangle}\ .
\end{equation}

With the fourth power we can construct another interesting quantity,
the cumulant $g_4$, defined as
\begin{equation}
g_4=\frac{3}{2}-\frac{1}{2}\frac{\overline{\langle {\cal M}^4\rangle}}
           {\overline{\langle {\cal M}^2 \rangle}^2}\ .
\end{equation}
In the finite-size scaling method we use it is very convenient to have
a well behaved estimate of the correlation length in a finite lattice. 
We have used the second-momentum definition, that reads~\cite{COOPER}
\begin{equation}
\xi=\left(\frac{\chi/F-1}{4\sin^2(\pi/L)}\right)^\frac{1}{2},
\label{XI}
\end{equation}
where $F$ is defined in terms of the Fourier transform of the
spin distribution
\begin{equation}
{\cal G}(\mbox{\boldmath$p$})=
\frac{1}{V}\sum_{\mbox{\boldmath\scriptsize$r$}}{\mbox{e}}^
{\mbox{\scriptsize i}
\mbox{\boldmath\scriptsize$p$}\cdot\mbox{\boldmath\scriptsize$r$}}
\epsilon_{\mbox{\boldmath\scriptsize$r$}}
\sigma_{\mbox{\boldmath\scriptsize$r$}}\ ,
\end{equation}
as
\begin{equation}
F=\frac{V}{3}\overline{\left\langle |{\cal
G}(2\pi/L,0,0)|^2+\mbox{permutations}\right\rangle}\ .
\end{equation}

\section{Finite-size scaling techniques}

In the scaling region, the mean value of a given  observable, $O$, 
measured at a coupling ($\beta , p$) pair can be written as 
\begin{equation}
O(L,\beta,p)=L^{x_O/\nu}\left(F_O(\xi(L,\beta,p)/L)
        +{\cal O}(L^{-\omega})\right)\ ,
\label{FSSEQ}
\end{equation}
where $x_O$ is the critical exponent of the operator $O$, $F_O$ is a 
smooth scaling function depending on the observable
and $\omega$ corresponds to the eigenvalue of the
first irrelevant operator of the theory from the Renormalization Group 
point of view.

The principal feature of Eq.~(\ref{FSSEQ}) is that all the quantities
are measurable in a finite lattice. In order to obtain the critical
exponents we need to remove the unknown scaling function $F_O$. Let us
define the quotient of a given observable $O$ at two different lattice
sizes and at the same coupling pair as
\begin{equation}
Q_O=O(sL,\beta,p)/O(L,\beta,p)\ ,
\end{equation}
and let us compute this quotient
at the coupling where the correlation length in units of the lattice 
size is the same for both lattices. Thus we get: 
\begin{equation}
\left.Q_O\right|_{Q_\xi=s}=s^{x_O/\nu}+A^O_p \, L^{-\omega} + \cdots
\label{QUOMEGA}\ ,
\end{equation}
where $A^O_p$ is a constant which depends on the observable and the 
spin concentration $p$ and the dots stand for higher-order scaling 
corrections. From this equation we can extract the critical
exponent associated to a given observable.

The observables used to obtain the different critical exponents are:
the $\beta$-derivative of the correlation length in order to calculate
$\nu$  ($x_{\partial_\beta \xi}=\nu+1$) and the
susceptibility, $\chi$, to get the magnetic $\eta$ exponent,
($x_{\chi}=\nu(2-\eta)$).

In order to compute the infinite volume critical coupling we will use
the crossing points of the observables with $x_O=0$, as $g_4$ or
$\xi/L$, when measured at two different lattice sizes, $L$ and $sL$.
The shift of these points from the critical coupling behaves
as~\cite{BINDER}:
\begin{equation}
\Delta \beta_{\rm c}^L  
\propto \frac{1-s^{-\omega}}{s^{1/\nu}-1}L^{-\omega-1/\nu}\ .
\label{BETACFIT}
\end{equation}

\section{Numerical methods}

The best update method for an Ising model simulation is a cluster
algorithm~\cite{SW}. In particular, the most efficient one in the pure
case is the single-cluster Wolff method~\cite{WOLFF}.  Nevertheless,
in a diluted system small groups of isolated spins appear, which are
scarcely visited with this algorithm.  Furthermore in the non-Gaussian
case also appear isolated occupied lines.  In order to update
all-sized spin clusters, after a fixed number of single-cluster
updates we perform a Swendsen-Wang sweep. We call this ensemble our MC
Step (MCS). We have discarded 100 MCS for thermalization and then we
have measured the different observables for every single MCS.  We have
checked the correct thermalization of the system by starting from hot
and cold configurations.  We have chosen the single-cluster updates
number in such a way that the autocorrelation times for all the
observables are nearly one MCS. The simulations are carried out in
the RTNN machine at Zaragoza University.

Other interesting parameters are the number of measures to perform in a
given disorder realization, $N_I$, using the Ising Hamiltonian, 
Eq.~(\ref{HAMILTONIAN}),  and the number of different samples, $N_S$.
We refer to~\cite{ISDIL3D,ISDIL2D4D} for a discussion of the optimal 
choice of these parameters. In our case, we have performed $N_I=100$ 
measures in  $N_S=20000$ different samples for $L\leq64$ and 
in $N_S=10000$ samples for $L=128$. 

We have used in this study the usual $\beta$-extrapolation
~\cite{FALCIONI}. Thus we restrict to not too strong dilutions.

In order to work with large dilutions it is convenient to perform a
$p$ extrapolation as we will see later from the phase diagram of the
systems.  In the random site-diluted Ising model case this is possible
because the density distribution probability  of the actual
configurations (a binomial one) is known.  In the Gaussian model, due
to the correlation between the different sites, this distribution it
is not known. Neither in the non-Gaussian case, as it is presented
here.  Nevertheless it is possible to perform a slight variation of
this latter model allowing to perform a dilution extrapolation. It is
enough to choose with a given probability when a line it is empty
or filled, but this variation will be not considered in this study.

We recall~\cite{ISDIL3D,ISDIL2D4D} that a bias of order
${2\tau}/{N_I}$ is present in the $\beta$-extrapolation, where $\tau$
is the correlation time between the energy and the observable we
consider. This fact is not relevant in the usual MC calculations,
because the statistical errors are of order $1/\sqrt{N_I}$.  But in
diluted system investigations, when $\sqrt{N_S}\sim N_I$, this bias
could be not negligible. We have performed a proper extrapolation
procedure~\cite{ISDIL3D,ISDIL2D4D} in order to obtain unbiased
estimates of the $\beta$-derivatives and the values of the different
observables in the neighbourhood of the simulated couplings. The MCS
is chosen in such a way that $\tau$ is nearly one measure.  For the
largest lattice we have considered, $L=128$, in the non-Gaussian case,
the single cluster update number for every Swendsen-Wang sweep is 1200
and for the Gaussian case it is 400.  For the statistical error
computation we have used the jack-knife method with 50 blocks, that
allows us to obtain a 10\% of accuracy in the error bars.

\section{Numerical results}

We have studied the Gaussian case at two different dilutions, $p=0.8$
and $p=0.65$, performing simulations in lattice sizes $L=8,16,32,64$
and 128. In the non-Gaussian case we have only considered $p=0.8$ in
the same lattice sizes.

\begin{figure}[t]
\begin{center}
\epsfig{file=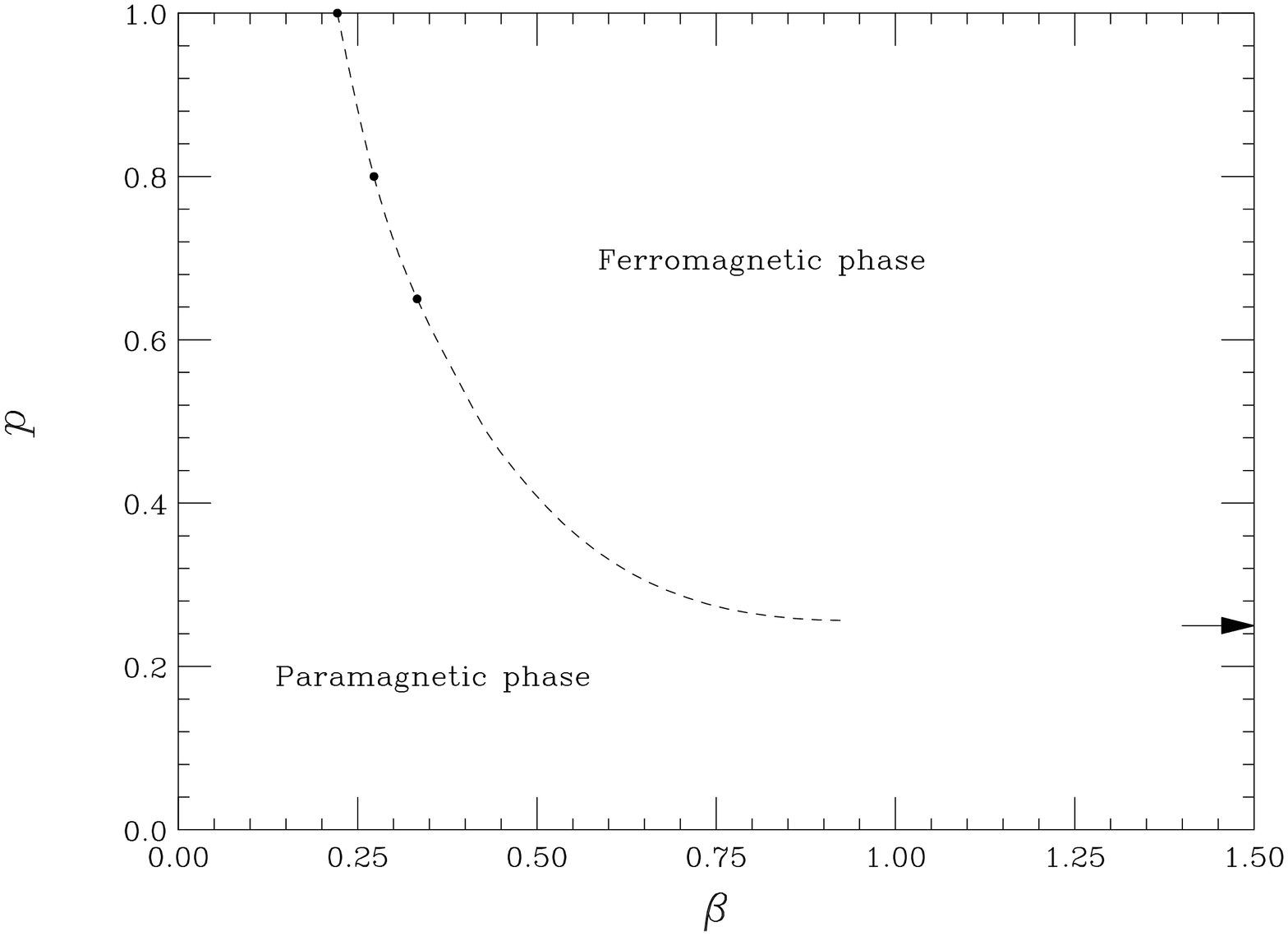,width=0.7\linewidth,angle=90}
\end{center}
\caption{Phase diagram of the site-diluted Ising model with long-range 
correlated Gaussian disorder, in the
inverse temperature--dilution plane. The dots correspond to the simulated
points, while the arrow points  to the percolation
limit ($\beta=\infty$).}
\label{DIAGRAM}
\end{figure}

In figure~\ref{DIAGRAM} we show the phase diagram for the Gaussian
model.  The percolation critical point, $p_{\rm c}\simeq 0.25$, was
obtained by studying the behaviour of the $g_4$ function in a $L=128$
lattice. In the thermodynamical limit, $g_4=0$ in the disordered phase
and $g_4=1$ in a ferromagnetic ordered one.  The corresponding phase
diagram for the non-Gaussian case is qualitatively the same, with a
ferromagnetic ordered phase for large $\beta$, when $p$ is larger that
the percolation threshold for this case.

\subsection{Thermal exponent}

In table~\ref{NUTABLE} we present the results for the $\nu$ 
exponent in the two cases considered, the Gaussian  and the non-Gaussian 
disorder. It was computed by applying the Eq.~(\ref{QUOMEGA}) to  
$\partial_\beta \xi$ using $s=2$.

As we see, there are visible scaling corrections in all the cases.
Nevertheless the values we obtain for the $\nu$ exponent are very
different from those of the pure Ising model,
$\nu=0.6294(5)(5)$~\cite{ISINGPERC} and from those of the
three-dimensional random site-diluted case,
$\nu=0.6837(24)(29)$~\cite{ISDIL3D}.  In order to obtain the critical
exponent, we have to perform an infinite volume  extrapolation
procedure.  It is possible that corrections-to-scaling that we
observe are complicated by the presence of oscillatory terms, as is
suggested by the first order of the $\epsilon$-expansion, but we are
unable to confirm or discard this possibility. Strong corrections
to simple scaling are present as we will see, so it is
rather difficult (although we study lattices ranging from $8^3$ to
$128^3$) to get conclusive statements on the nature of finite
volume corrections.

We can try to parameterize the scaling corrections as in
Eq.~(\ref{QUOMEGA}) only with the first term. We have used the data
from the two different dilutions of the Gaussian case performing a
joint fit assuming a single value for $\omega$ and $\nu$ exponents,
following the picture of a single Universality Class along the
critical line.  Using $L\geq 8$ data and the full covariance matrix to
compute the statistical function $\chi^2$, we find a very large value
of $\chi^2/{\rm d.o.f.}=13.9/4$. Nevertheless, discarding the data
from the $L=8,16$ pair, we find $\chi^2/{\rm d.o.f.}=1.20/2$. The
value obtained for the thermal exponent, $\nu=1.012(16)$, is
compatible with the analytic predictions. We also find
$\omega=1.01(13)$.

We can control the presence of the higher-order corrections in a
simple and na\"{\i}ve way. We could perform a quadratic fit for each dilution
with $L\geq 8$ data, assuming the $\omega=1$ value compatible with our 
results and only using the diagonal part of the covariance matrix.  If
we do so, we obtain for $p=0.8$, $\chi^2/{\rm d.o.f.}=0.76/1$ with
$\nu=1.012(10)$ and for $p=0.65$, $\chi^2/{\rm d.o.f.}=0.73/1$
and $\nu=1.005(14)$. So, the presence of second-order corrections
for the thermal exponent data seems reasonable. Furthermore we have found 
that the  $\nu$ value is not affected by the presence of these terms.

In the non-Gaussian case large finite volume corrections are also
present. Nevertheless we find that the estimates from the two biggest
lattice pairs for the $\nu$ exponent are compatible with the
analytical calculations.

\subsection{Magnetic exponent}

In table~\ref{ETATABLE} we present the estimates of the magnetic exponent
$\eta$ using the Eq.~(\ref{QUOMEGA}), from the susceptibility $\chi$ 
measured at the point where $Q_\xi=2$ for all the concentrations 
considered.

As we can see from the table, there are strong scaling effects in all
the cases, specially in the non-Gaussian case. An infinite volume
extrapolation procedure is therefore 
needed in order to get an $\eta$ estimate.

Assuming only the presence of first-order corrections with our
previously calculated $\omega$ value we do not find reasonable fits to
our data.  So, we could consider higher-order correction terms. As we
have found $\omega\simeq 1$, the second-order terms and the analytical
corrections are of the same order, so we can try a quadratic joint fit
using the $\omega=1$ value. With the $L\geq 16$ data for all the
concentrations studied, using only the diagonal part of the
correlation matrix, we get $\chi^2/{\rm d.o.f.}=1.63/2$ and
$\eta=0.043(4)$. So we have found compatible results with the picture
of a single $\eta$ value, scaling corrections parameterized by
$\omega\simeq 1$, but with non-negligible higher-order correction
effects.  Nevertheless, this estimate has two different sources of
systematic error: the first one is due to the possible evolution of
the $\eta$ value with the minimum lattice size considered in the fits,
and the second, to the uncertainty on the fitted functional form.

We can compare this result with those from the random
site-diluted Ising model, $\eta=0.0374(36)(9)$~\cite{ISDIL3D},
and with those from the Ising case, $\eta=0.0374(6)(6)$~\cite{ISINGPERC},
finding that they are similar to our estimate for $\eta$.

\subsection{Critical couplings}

In table~\ref{CROSSINGTABLE} we show the crossing points of $g_4$ and
$\xi/L$ from ($L$,$2L$) lattice pairs for the Gaussian case. 
As we can see in the table, there is a non monotonic $L$ behaviour
for $g_4$ crossing points in both concentrations, so it is expected the
presence of high-order scaling corrections. 
A way to extract the infinite volume critical coupling is 
to perform a fit to the functional form of
the Eq.~(\ref{BETACFIT}).  In order to find a
proper extrapolated value, we have to check that we are in the linear
regime and that we can control the higher-order corrections.  
As the crossing points for $g_4$ show a minimum value around the $L=32,64$ 
pair the former condition is not satisfied.

We could fit the $\xi/L$ crossing points to Eq.~(\ref{BETACFIT}).
Nevertheless, for both concentrations, using $L\geq 8$ data, we have
found a large value of $\chi^2/{\rm d.o.f.}$, being d.o.f.=1. So we
have also to assume the presence of higher-order scaling corrections.

In order to control the finite volume effects, assuming our estimate
for $\omega+1/\nu=2.00(13)$, we can discard the $L=8$ data and perform
a linear fit for the $\xi/L$ data. In the $p=0.65$ case we find a
reasonable $\chi^2/{\rm d.o.f.}=1.28/1$ in the central value, and we
get $\beta_{\rm c}(\infty)=0.332929(13)(12)$, being the second error
bar due to the uncertainty in $\omega+1/\nu$. Nevertheless, for
$p=0.8$ we do not find a reasonable fit, showing that the higher-order
corrections are important also in the $L=16$ lattice. We can check
this latter picture performing a fit with $1/L^2$ and $1/L^3$ terms,
only using the diagonal part of the covariance matrix for $L\geq 8$
data. Then we obtain $\chi^2/{\rm d.o.f.}=1.69/1$ and $\beta_{\rm
c}(\infty)=0.272715(10)$. So the picture of second-order scaling
corrections is compatible with our data in this case.

A similar analysis can be done by studying the $g_4$ and $\xi/L$
crossing points measured with a ($L_1, L_2$) pair but fixing the $L_1$
value. 

In the $p=0.65$ case performing a
linear fit for $\xi/L$ with $L\geq 16$, and $\omega+1/\nu=2.00(13)$ we
get $\chi^2/{\rm d.o.f.}= 1.43/1$ for the central value of this
interval and $\beta_{\rm c}(\infty)=0.332927(13)(15)$, where the
second error bar is due to the uncertainty on the critical exponents.

In the $p=0.8$ case a diagonal fit for $\xi/L$ with $1/L^2$ and
$1/L^3$ terms using $L\geq 8$ data, we get $\chi^2/{\rm d.o.f.}=0.52/1$
and $\beta_{\rm c}(\infty)=0.272722(10)$.  So we find a similar
behaviour with our previous analysis, finding reasonable fits and
compatible estimates for the infinite volume critical couplings.

In  table~\ref{CROSSINGLINES} we show the crossing points of $g_4$ 
and $\xi/L$ measured at $L$ and $2L$ lattice sizes for the non-Gaussian 
case with mean concentration $p=0.8$.

Also in this case we see that the $g_4$ crossing point is not a
monotonic function of $L$.  In the $\xi/L$ case we find that assuming
our previous $\omega$ estimate, a linear fit for $L\geq 16$ is not
reasonable, so we have to conclude that also in this case the
higher-order terms are present. In order to check in a simple way this
assumption, we perform a fit with $1/L^2$ and $1/L^3$ terms, using
$L\geq 8$ and using only the diagonal part of the covariance
matrix. Thus we obtain $\chi^2/{\rm d.o.f.}=0.34/1$ and $\beta_{\rm
c}(\infty)=0.257126(14)$, so this picture is compatible with our data.

In order to compute the value of the scaling functions $g_4$
and $\xi/L$ at the critical coupling in the thermodynamical limit, we
have measured the values of these quantities at the crossing points of $g_4$
and $\xi/L$ respectively. In the $g_4$ case, the finite volume corrections
are large, finding values for this observable in the range 0.58-0.64. In
the $\xi/L$ case, we have also found that an infinite volume extrapolation
procedure is needed. Performing a $1/L$ extrapolation we quote for 
this quantity the value 0.36(2).

\section{Conclusions}

We have studied by Monte Carlo means the three-dimensional
site-diluted Ising model, with long-range spatially correlated
disorder. We have considered Gaussian and non-Gaussian disorder 
to study the influence of this fact in the critical behaviour of
the system.  We have used finite-size scaling techniques for 
the computation of the critical exponents.

We have found strong scaling-corrections for the $\nu$ exponent.  In
the Gaussian case, we succeed to parameterize them only with the first
corrections-to-scaling term, finding an infinite volume $\nu$ value
that is compatible with the analytical prediction in this model.  In
the non-Gaussian case, the value we obtain for the two largest
lattice-size pairs is also compatible with this calculation.

For $\eta$ exponent, large finite volume effects are also present.
Our data for $L\geq 16$ are compatible with the picture of a single
$\eta$ value independent of the kind of disorder and the concentration
considered, but with non-negligible second-order scaling-correction
terms.

So we have obtained a consistent picture of the existence of a single
fixed point (single $\eta$, $\nu$ and $\omega$ values) using 
Gaussian and non-Gaussian correlated disorder, but with non-negligible
second-order scaling-corrections. This fact introduces 
systematic errors in our analysis making it very difficult to measure them, 
in order to obtain solid estimates for the critical exponents.

\section{Acknowledgments}
We thank the CICyT (contracts AEN97-1693 and AEN97-1708) for partial
financial support.  We also thank the RTNN collaboration specially for
computing facilities. H.~G.~B.  is grateful to the department of
Physics of the University of Rome I, {\sl La Sapienza}, for the
hospitality during the completion of this work. H.~G.~B. is partially
supported by the U.C.M.  We are grateful to M. Altarelli for pointing
this interesting problem to us and for an illuminating discussion.

\begin{table}
\caption{Critical exponent $\nu$ 
computed using $\partial_\beta \xi$ when  
measured in $(L,2L)$ lattice pairs at the couplings where $Q_\xi=2$  
for both models at the different concentrations considered.}
\begin{center}
\begin{tabular}{rllcl}
\multicolumn{1}{c} {}  & 
\multicolumn{2}{c}{Gaussian model} &
\multicolumn{1}{c} {}  &
\multicolumn{1}{c}{Non-Gaussian} \\\cline{2-3}\cline{5-5}
\multicolumn{1}{c} {$L$}  & 
\multicolumn{1}{c}{$p=0.8$} &
\multicolumn{1}{c}{$p=0.65$} &
\multicolumn{1}{c} {}  & 
\multicolumn{1}{c}{$p=0.8$} \\\hline
8 &   0.7626(19)&  0.871(3)  &&  0.8335(24)\\
16&   0.833(3)&  0.942(6)  &&  0.934(4)\\
32&   0.907(4)&  0.969(7)  &&  1.009(9)\\
64&   0.964(9)&  0.996(11)   &&  1.009(13)\\
\end{tabular}
\label{NUTABLE}
\end{center}
\end{table}

\begin{table}[t]
\caption{Magnetic exponent $\eta$ computed  from $\chi$, using lattice  
($L,2L$) pairs at the couplings where  $Q_\xi=2$ for both Gaussian and 
non-Gaussian cases at the different concentrations simulated.}
\begin{center}
\begin{tabular}{rllcr}
\multicolumn{1}{c} {}  & 
\multicolumn{2}{c}{Gaussian model} &
\multicolumn{1}{c} {}  &
\multicolumn{1}{c}{Non-Gaussian} \\\cline{2-3}\cline{5-5}
\multicolumn{1}{c} {$L$}  & 
\multicolumn{1}{c}{$p=0.8$} &
\multicolumn{1}{c}{$p=0.65$} &
\multicolumn{1}{c} {}  & 
\multicolumn{1}{c}{$p=0.8$} \\\hline
8 &        0.0085(11)  &  0.0256(14) &&  -0.0513(14) \\
16&        0.0082(14)  &  0.0274(16) &&  -0.0532(12) \\
32&        0.0137(15)  &  0.0384(18) &&  -0.0259(18) \\
64&        0.0259(19)  &  0.038(3)   &&   0.0052(24)\\
\end{tabular}
\label{ETATABLE}
\end{center}
\end{table}

\begin{table}
\caption{Crossing points from ($L,2L$) pairs  of 
$g_4$ and $\xi/L$  for the Gaussian case at the different concentrations 
simulated.}
\begin{center}
\begin{tabular}{rllcll}
\multicolumn{1}{c} {}  & 
\multicolumn{2}{c}{$p=0.8$} &
\multicolumn{1}{c}{} &
\multicolumn{2}{c}{$p=0.65$}  \\\cline{2-3}\cline{5-6}
\multicolumn{1}{l}{$L$} & 
\multicolumn{1}{c}{$\xi/L$}&
\multicolumn{1}{c}{$g_4$}&
\multicolumn{1}{c}{} &
\multicolumn{1}{c}{$\xi/L$}&\multicolumn{1}{c}{$g_4$}
\\\hline
8 &0.274535(34) & 0.273760(52) &&0.335269(76)&  0.33358(12) \\
16&0.273545(15) & 0.272862(22) &&0.333709(41)&  0.332617(72)\\
32&0.2729883(96)& 0.272604(14) &&0.333099(19)&  0.332682(28)\\
64&0.2727805(70)& 0.272624(11) &&0.332989(15)&  0.332872(25)\\
\end{tabular}
\label{CROSSINGTABLE}
\end{center}
\end{table}

\begin{table}
\caption{Crossing points of $g_4$ and $\xi/L$ from $(L,2L)$ pairs for $p=0.8$ 
in the non-Gausssian disorder.}
\begin{center}
\begin{tabular}{rll}
\multicolumn{1}{l}{$L$} & \multicolumn{1}{c}{$\xi/L$}&
\multicolumn{1}{c}{$g_4$}\\\hline
8 & 0.25926(4)   &0.25803(5)  \\
16& 0.257935(22) &0.25706(3)  \\
32& 0.257375(13) &0.25708(21)  \\
64& 0.257188(9)  &0.257110(13) \\
\end{tabular}
\label{CROSSINGLINES}
\end{center}
\end{table}

\end{document}